\documentstyle[12pt]{article}
\font\goth=eufm10
\font\double=msym10
\font\ccal=cmsy8

\def\cc{{\hbox{\double C}}}

\def\rr{{\hbox{\double R}}}
\def\zz{{\hbox{\double Z}}}
\def\qqq{{\hbox{\double Q}}}
\def\aa{{\cal A}}
\def\dd{{\cal D}}
\def\gg{\hbox{\goth g}}
\def\hh{{\cal H}}
\def\hhh{{\hbox{\double H}}}
\def\ff{{\cal F}}
\def\ll{{\cal L}}
\def\mm{{\hbox{\ccal M}}}
\def\sss{{\cal S}}
\def\t{\,{\rm tr}\,}

\def\ddd{{\,\hbox{$\partial\!\!\!/$}}}
\def\dee{\,\hbox{\rm D}}
\def\de{\,\hbox{\rm d}}

\def\pa{{\partial}}

\def\lb{\left[}
\def\rb{\right]}
\def\lp{\left(}
\def\rp{\right)}

\def\ul{\underline}
\def\ot{\otimes}
\def\op{\oplus}

\def\bb{\begin{eqnarray}}
\def\ee{\end{eqnarray}}
\def\eee{\nonumber\end{eqnarray}}
\def\pp{\pmatrix}
\def\qq{\quad}

\begin{document}

\hsize 17truecm
\vsize 24truecm
\font\twelve=cmbx10 at 13pt
\font\eightrm=cmr8
\baselineskip 18pt

\begin{titlepage}

\centerline{\twelve CENTRE DE PHYSIQUE THEORIQUE}
\centerline{\twelve CNRS - Luminy, Case 907}
\centerline{\twelve 13288 Marseille Cedex}
\vskip 4truecm

\centerline{\twelve YANG-MILLS-HIGGS
versus CONNES-LOTT}

\bigskip

\begin{center}
{\bf Bruno IOCHUM}
\footnote{ and Universit\'e de Provence} \\
\bf Thomas SCH\"UCKER $^{1}$
\end{center}

\vskip 2truecm
\leftskip=1cm
\rightskip=1cm
\centerline{\bf Abstract}

\medskip

By a suitable choice of variables we show that every
Connes-Lott model is a Yang-Mills-Higgs model. The
contrary is far  from being true. Necessary conditions
are given. Our analysis is pedestrian and illustrated by
examples.

\vskip 2truecm
PACS-92: 11.15 Gauge field theories\\
\indent
MSC-91: 81E13 Yang-Mills and other gauge theories

\vskip 3truecm

\noindent dec. 1994
\vskip 1truecm
\noindent CPT-94/P.3090\\
hep-th/9501142

\vskip 3truecm

 \end{titlepage}

Despite its impressing success in describing particles
and interactions, the Yang-Mills-Higgs (YMH) model
building kit has conceptual short comings:
\begin{itemize}
\item
 its rules are essentially unmotivated,
\item
 its complicated input comprising a Lie group and three
representations is only justified by experiment,
\item
the model singled  out by more and more
precise experiments, namely the standard
$SU(3)\times SU(2)\times U(1)$ model of electro-weak
and strong interactions, is ugly and nobody really
believes it to be the last word.
\end{itemize}
 Concerning the first two points, the
Connes-Lott (CL) model building kit \cite{cl} does
better. Its rules have a precise motivation from
non-commutative geometry and its input, comprising
an involution algebra and two representations, is
infinitely more restricted than the YMH input.
Nevertheless, the standard model is also a CL model
\cite{cl,k,ks,vg}, a fact that by itself does not improve
its beauty, but that perhaps allows unification with
gravity. Indeed, the Einstein-Hilbert action as well may
be formulated naturally in the setting of
non-commutative geometry \cite{cG,kG,kw}.

The purpose of this work is to show that the CL models
represent a very small subset of the YMH models,
where we restrict ourselves to ``local'' models, i.e.
models defined on trivial bundles. Also we restrict
ourselves to CL models defined by means of a finite
dimensional algebra $\aa$ tensorized with the algebra
of functions on (a compact, Euclidean) ``spacetime'' of
dimension 4.
 These
particular models can be computed with elementary
mathematics \cite{sz} and compare naturally to YMH
models. Models whose algebras are not such
tensor products, as the non-commutative torus
\cite{cr}, the fuzzy sphere \cite{m} or a quantum
space time \cite{dfr}
 are much more involved
mathematically and appear as natural candidates for
the above mentioned unification.

\section{ Yang-Mills-Higgs models}

Let us first set up our notations of a YMH model. It is
defined by the following input:
\begin{itemize}
\item
 a finite dimensional, real, compact Lie group $G$,
\item
 a positive definite, bilinear
invariant form on the Lie algebra  $\gg$ of $G$, this
choice being parameterized by a few positive numbers
$g_i$, the coupling constants,
\item
a (unitary) representation $\rho_L$ on a Hilbert space
$\hh_L$ accommodating the left handed fermions
$\psi_L$,
  \item
a representation $\rho_R$ on $\hh_R$ for the right
handed fermions $\psi_R$,
\item
 a representation $\rho_S$ on $\hh_S$ for the scalars
$\varphi$,  \item
 an invariant,
positive polynomial $V(\varphi),\  \varphi \in \hh_S$
 of order 4, the Higgs potential,
\item
 one complex number or Yukawa
coupling $g_Y$ for every trilinear invariant, i.e. for
every one dimensional invariant subspace, ``singlet'', in
the decomposition of the representation  associated to
$  \left(\hh_L^{\ast}\otimes
\hh_R\otimes \hh_S\right) \oplus\left(
     \hh_L^\ast\otimes \hh_R\otimes \hh_S^*\right).
$
\end{itemize}
The standard model is defined by the following input:
\bb G =  SU(3) \times SU(2) \times U(1) \eee
with three coupling constants $ g_3, g_2, g_1, $
\bb
\hh_L &=& \bigoplus_1^3\lb (1,2,-1)\oplus (3,2,{1
\over 3}) \rb  \label{HL},\\
 \hh_R& = &\bigoplus_1^3\lb (1,1,-2)\oplus
(3,1,{4\over 3})\oplus (3,1,-{2\over 3}) \rb, \cr\cr
 \hh_S &= &(1,2,-1) ,\eee
where $(n_3,n_2,y)$
denotes the tensor product of an $n_3$ dimensional
representation of $SU(3)$, an $n_2$ dimensional
representation of $SU(2)$ and the one dimensional
representation of $U(1)$ with hypercharge $y$:
\bb
\rho(e^{i\theta}) = e^{iy\theta}, &&\qquad
y\in\qqq,\   \theta \in [0,2\pi),\cr
 V(\varphi) = \lambda (\varphi^\ast
\varphi)^2 - {{\mu ^2 }\over 2}\varphi^\ast \varphi,
&&\qquad \varphi \in
\hh_S, \quad \lambda, \mu > 0.\eee
There are 27 Yukawa couplings in the standard model.

The gauge symmetry is said to be spontaneously broken
 if every minimum $v\in \hh_S$ of the
Higgs potential is gauge variant,
$ \rho_S(g)v\ne v$ for some $ g\in G$.
Any such minimum $v$ is called a vacuum. For
simplicity let us assume that the vacuum is
non-degenerate, i.e. all minima lie on the same orbit
under $G$. Then, the little group $G_\ell$ is by
definition the isotropy group of the vacuum,
$\rho(g_\ell)v=v$. For example, in the standard model
any doublet of length
$\sqrt{\varphi^*\varphi}=\mu/(2\sqrt{\lambda})$ is a
minimum and the little group is
$ G_\ell=SU(3)\times U(1)_{em}.$
To do perturbation theory, we have to introduce a scalar
variable $h$, that vanishes in the vacuum,
\bb h(x):=\varphi(x)-v,\eee
$x$ a point in spacetime $M$.
With this change of variables, the Klein-Gordon
Lagrangian is $ \left({\rm
D}\varphi\right)^**{\rm D} \varphi$.
 The Hodge star $*\cdot$
should be distinguished from the Hilbert space dual
$\cdot^*$, wedge symbols are suppressed.
 We denote by D
the covariant exterior derivative, here for scalars
$
\dee\varphi:=\de\varphi+\tilde\rho_S(A)\varphi,$
$\varphi$ is now a multiplet of {\it fields}, i.e. a 0-form
on spacetime with values in the scalar representation
space,
$ \varphi\in\Omega^0(M,\hh_S),$
while the vacuum $v$ remains  constant over
spacetime so that it also minimizes the kinetic term
$\de\varphi^**\de\varphi$. The gauge fields are
1-forms with values in the Lie algebra of $G$:
$ A\in\Omega^1(M,\gg),$
$\tilde\rho_S$ denotes the Lie algebra representation
on $\hh_S$. The Klein-Gordon Lagrangian produces
the mass matrix for the gauge bosons $A$.
This mass matrix is given by the (constant) symmetric,
positive semi-definite form on the Lie algebra of $G$,
\bb \left(\tilde\rho_S(A)v\right)^*\tilde\rho_S(A)v.
\eee
It contains the masses of the gauge bosons and
vanishes on the generators of the little group. In the
example of the standard model, the little group is
generated by the gluons and the
photon which remain massless.

In the following we are more concerned with the
fermionic mass matrix $\mm$, a linear map $ \mm:
\hh_R \longrightarrow \hh_L$. We want to produce it
in the same way we produced the mass matrix for the
gauge bosons, via the change of variables
$h(x):=\varphi(x)-v$. For this purpose, we add by hand
to the Dirac Lagrangian gauge invariant trilinears
\bb \sum_{j=1}^ng_{Yj}\left(\psi_L^*,\psi_R,\varphi
\right)_j+\sum_{j=n+1}^mg_{Yj}\left(\psi_L^*,\psi_R,
\varphi^*\right)_j\ +\ {\rm complex\ conjugate}
\label{tri},\ee
$n$ is the number of singlets in
$\left(\hh_L^{\ast}\otimes\hh_R\otimes \hh_S\right)
$, $m+n$ the number of singlets in
$\left(
     \hh_L^\ast\otimes \hh_R\otimes \hh_S^*\right)$.
For $h=0$ again, we
obtain the fermionic mass matrix $\mm$ as a function
of the Yukawa couplings $g_{Yj}$ and the vacuum $v$
\bb \psi_L^*\mm\psi_R:         =
\sum_{j=1}^ng_{Yj}\left(\psi_L^*,\psi_R,v
\right)_j+\sum_{j=n+1}^mg_{Yj}\left(\psi_L^*,\psi_R,
v^*\right)_j.\eee
 As
the gauge boson masses, the fermionic mass terms
$\psi_L^*\mm\psi_R$ are not gauge invariant in
general. They are gauge invariant if
$ \rho_L(g^{-1})\mm\rho_R(g)=\mm\
{\rm for\ all}\ g\in G$
and in analogy with the little group, we define
$G_\mm$ to be the subgroup of $G$, that leaves $\mm$
invariant,
\bb
\rho_L(g_\mm^{-1})\mm\rho_R(g_\mm)=\mm\quad
{\rm for\ all}\ g_\mm\in G_\mm.\eee
In the standard model with its 27 Yukawa couplings,
the mass matrix $\mm$ can be any matrix yielding
mass terms invariant under the little group.

In general however, the little group is only a subgroup
of $G_\mm$,
\bb G_\ell\subset G_\mm.\eee
For example, if we modify the standard model by
choosing the scalars in the adjoint representation of
$SU(2)$ then the little group becomes
$ G_\ell= SU(3)\times U(1)\times U(1)$,
there are no trilinear invariants, the mass matrix
$\mm$ vanishes, and
$G_\mm=SU(3)\times SU(2)\times U(1).$

\section{ Connes-Lott models}

With the two specializations mentioned in the
introduction, a Connes-Lott model is defined by the
following choices:
\begin{itemize}
\item
 a finite dimensional, associative,
algebra $\aa$ over the field $\rr$ or $\cc$ with unit 1
and involution  \nolinebreak $^*$,
\item
 two *- representations of $\aa$, $\rho_L$ and
$\rho_R$, on Hilbert spaces $\hh_L$ and $\hh_R$
over the field, such
that $\rho:=\rho_L\op\rho_R$ is faithful,
\item
 a mass matrix $\mm$ i.e. a linear map
$ \mm: \hh_R \longrightarrow\hh_L, $
\item
a certain number of
coupling constants depending on the degree of
reducibility of $ \rho_L\oplus \rho_R. $
\end{itemize}
The data $(\hh_L,\hh_R,\mm)$ plays
a fundamental role in non-commutative Riemannian
geometry where it is called K-cycle.

With this input and the rules of non-commutatative
geometry, Connes and Lott construct a YMH model.
Their starting point is
an auxiliary differential algebra $\Omega \aa,$
the so called universal differential envelope of $\aa$:
\bb   \Omega^0\aa := \aa,\eee
$\Omega^1\aa$ is
generated by symbols $\delta a$, $ a \in \aa$ with
relations
\bb   \delta 1 = 0,\qq     \delta(ab) = (\delta
a)b+a\delta b.\eee
 Therefore $\Omega^1\aa$ consists
of finite sums of terms of the form $a_0\delta a_1,$
\bb   \Omega^1\aa = \left\{ \sum_j a^j_0\delta
a^j_1,\quad a^j_0, a^j_1\in \aa\right\}\eee
and likewise
for higher $p$,
\bb   \Omega^p\aa = \left\{ \sum_j
a^j_0\delta a^j_1...\delta a^j_p ,\quad a^j_q\in
\aa\right\}.\eee
 The differential $\delta$ is defined by
$   \delta(a_0\delta a_1...\delta a_p) :=
   \delta a_0\delta a_1...\delta a_p.$

Two remarks: The universal differential envelope
$\Omega\aa$ of a commutative algebra $\aa$ is
not necessarily graded commutative. The
universal differential envelope of any algebra has no
cohomology. This means that every closed form
$\omega$ of degree  $p\geq 1,\quad
\delta\omega=0,$ is exact, $\omega =
\delta\kappa$ for some $(p-1)$form $\kappa.$

The involution $^*$ is
extended from the algebra $\aa$ to
 $\Omega^1\aa$ by putting
\bb   (\delta a)^* := \delta(a^*) =:\delta a^*.\eee
Note that
Connes defines $(\delta a)^* := -\delta(a^*)$ which
amounts to replacing $\delta$ by $i\delta.$ With the
definition
$ (\omega\kappa)^*=\kappa^*\omega^*$,
the involution is extended to the whole
differential envelope.

The next step is to extend the
representation $\rho:=\rho_L\op\rho_R$ on
$\hh:=\hh_L\op\hh_R$ from the  algebra $\aa$   to its
universal differential envelope $\Omega\aa$. This
extension is the central piece of Connes' algorithm
and deserves a new name:
\bb && \pi :
\Omega\aa  \longrightarrow{\rm End}(\hh)
   \cr
 &&\pi(a_0\delta a_1...\delta a_p) :=
(-i)^p\rho(a_0)[\dd,\rho(a_1)] ...[\dd,\rho(a_p)]
\label{pi}\ee
where $\dd$ is the linear map from
$\hh$ into itself
\bb   {\dd}:= \pmatrix {0 & \mm \cr\mm^*
&0}.\eee
In non-commutative geometry, $\dd$ plays the role of
the Dirac operator and we call it internal
Dirac operator.
Note that in Connes' notations there is no factor
$(-i)^p$ on the rhs of equation (\ref{pi}). A
straightforward calculation shows that $\pi$ is in fact
a  representation of $\Omega\aa$ as involution
algebra, and we are tempted to define also a
differential, again denoted by $\delta$,
 on $\pi(\Omega\aa)$ by
\bb
\delta\pi(\omega):=\pi(\delta\omega).
\label{trial}\ee
However, this definition does not make sense if there
are forms
$\omega\in\Omega\aa$ with
$\pi(\omega)=0$ and  $\pi(\delta\omega)
\not= 0$. By dividing out these unpleasant forms,
Connes constructs a new differential algebra
$\Omega_\dd\aa$, the  interesting object:
\bb   \Omega_\dd\aa :=
{{\pi\left(\Omega\aa\right)}\over J}\eee
with
\bb   J := \pi\left(\delta\ker\pi\right) =:
\bigoplus_p J^p\eee
($J$ for junk).
On the quotient now,
the differential (\ref{trial}) is well defined. Degree by
degree we have:  \bb   \Omega_\dd^0\aa =\rho(\aa)\eee
because $J^0=0$ ,
\bb   \Omega_\dd^1\aa = \pi(\Omega^1\aa)\eee
because $\rho$ is faithful,
and in degree $p\geq2$,
\bb   \Omega_\dd^p\aa =
{{\pi(\Omega^p\aa)}\over
{\pi(\delta(\ker\pi)^{p-1})}}.\eee
While $\Omega\aa$ has no cohomology,
$\Omega_\dd\aa$ does in general. In fact, in infinite
dimensions,
if $\ff$ is the algebra of complex functions on
spacetime $M$ and if the K-cycle is
obtained from the Dirac operator then
$\Omega_\ddd\ff$ is de Rham's differential algebra of
differential forms on $M$.

We come back to our finite
dimensional case. Remember that the elements of the
auxiliary differential algebra $\Omega\aa$ that
we introduced for book keeping purposes only, are
abstract entities defined in terms of symbols and
relations. On the other hand the elements of
$\Omega_\dd\aa$, the ``forms'', are operators on the Hilbert
space $\hh$, i.e. concrete matrices of
complex numbers. Therefore there is a
natural scalar product defined by
\bb   <\hat\omega,\hat\kappa> :=
\t (\hat\omega^*\hat\kappa),
\quad  \hat\omega, \hat\kappa \in
\pi(\Omega^p\aa)\label{sp}\ee
for elements of equal degree and by zero for two
elements of different degree.
With this scalar product  $\Omega_\dd\aa$ is a
subspace of $\pi(\Omega\aa)$, by  definition
orthogonal to the junk. As a subspace,
$\Omega_\dd\aa$  inherits a scalar product which deserves
a special name ( , ). It is  given by
\bb   (\omega,\kappa) =
<\hat\omega,P\hat\kappa>, \quad \omega, \kappa
\in \Omega_\dd^p\aa\eee
 where $P$ is the orthogonal
projector in $\pi(\Omega\aa)$  onto the
ortho-complement of $J$ and $\hat\omega$ and
$\hat\kappa$ are any  representatives in the classes
$\omega$ and $\kappa$. Again the scalar product
vanishes  for forms with different degree. For real
algebras, all traces must be understood as real part of
the trace.

In Yang-Mills models coupling constants appear as
parameterization of the most general gauge invariant
scalar product. In the same spirit, we want the most
general scalar product on $\pi(\Omega\aa)$
compatible with the underlying algebraic structure. It
is given by
\bb   <\hat\omega,\hat\kappa>_z :=
\t (\hat\omega^*\hat\kappa\,z),
\quad  \hat\omega, \hat\kappa \in
\pi(\Omega^p\aa),\label{gsp}\ee
where $z$ is a positive operator on $\hh$, that
commutes with $\rho(\aa)$ and with the Dirac
operator $\dd$ and that leaves $\hh_L$ and $\hh_R$
invariant. A natural subclass of these scalar products
is constructed with operators $z$ in the image under
$\rho$ of the center of $\aa$.

Since $\pi$ is a homomorphism of involution algebras
the product  in $\Omega_\dd\aa$ is given by matrix
multiplication followed by the  projection $P$.
The involution  is
given by transposition and complex conjugation, i.e.
the dual with respect to the scalar product of the Hilbert
space $\hh$. Note that this scalar product admits no
generalization. W. Kalau et al. \cite{kppw} discuss the
computation of the junk and of the differential for
matrix algebras.

At this stage there is a first contact with gauge
theories.  Consider the vector space of anti-Hermitian
1-forms
$   \left\{ H\in \Omega_\dd^1\aa,\ H^*=-H
\right\}.$
A general element $H$ is of the form
\bb   H =
i\pmatrix {0&h\cr h^*&0}\eee
with $h$ a finite sum of terms
$\rho_L(a_0)[\rho_L(a_1)\mm-\mm\rho_R(a_1)]:\qq
\hh_R \rightarrow \hh_L\quad a_0,a_1\in\aa.$
 These elements are
called Higgses or gauge potentials.  In fact the space of
gauge potentials carries an affine  representation of
the group of unitaries
\bb    G: = \{g\in \aa,\ gg^\ast
=g^*g=1\}\eee
defined by
\bb H^g &:=&\
\rho(g)H\rho(g^{-1})+\rho(g)\delta (\rho(g^{-1})) \cr
       &=&\ \rho(g)H\rho(g^{-1})+(-i)\rho(g)[\dd,\rho
(g^{-1})] \cr
      &=&\ \rho(g)[H-i\dd]\rho(g^{-1})+i\dd\cr\cr
       &=& i\pmatrix {0&h^g\cr (h^g)^*&0}\eee
with
$    h^g-\mm :=
\rho_L(g)[h-\mm]\rho_R(g^{-1})$.
 $H^g$ is
the ``gauge transformed of $H$''.  As usual every gauge
potential $H$ defines a covariant derivative  $\delta
+H$, covariant under the left action of $G$ on
$\Omega_\dd\aa$:
\bb   ^g\omega := \rho(g)\omega, \quad
\omega\in\Omega_\dd\aa\eee
 which means
\bb   (\delta+H^g)\
^g\omega = \ ^g\lb(\delta+H)\omega\rb.\eee
 Also we define the
curvature $C$ of $H$ by
\bb   C := \delta H+H^2\ \in\Omega_\dd^2\aa.\eee
Note that here and later, $H^2$ is considered as element
of $\Omega_\dd^2\aa$ which means it is the projection $P$
applied to $H^2\in \pi(\Omega^2\aa)$.
The curvature $C$ is a Hermitian 2-form with {\it
homogeneous} gauge transformations
 \bb   C^g :=
\delta(H^g)+(H^g)^2 = \rho(g) C \rho(g^{-1}).\eee
Finally, we define the preliminary Higgs potential
$V_0(H)$, a functional on the  space of gauge
potentials, by
\bb   V_0(H)
:= (C,C) = \t[(\delta H+H^2)P(\delta H+H^2)].\eee
It is a
polynomial of degree 4 in $H$ with real, non-negative
values.  Furthermore it is gauge invariant,
$V_0(H^g)= V_0(H)$,
 because of the homogeneous transformation
property of the  curvature $C$ and because the
orthogonal projector $P$ commutes  with all gauge
transformations,
$\rho(g)P =P\rho(g)$.
The most remarkable property of the preliminary
Higgs potential is that, in most cases, its vacuum
spontaneously breaks the group $G$. To see this, define
\bb \dd_G:=-i\int_G\pi(g^{-1}\delta g)\de g\eee
where $\de g$ is the Haar measure of the compact Lie
group $G$. Thus $\dd_G$ is in $\Omega_\dd^1\aa$, unlike
the internal Dirac operator $\dd$ which is not
necessarily in $\Omega_\dd^1\aa$, see the next example.
Moreover
\bb \dd_G= \dd-\int_G\rho(g^{-1})\dd\rho(g)\de g
=\pp{0&\mm_G\cr \mm_G^*&0}\eee
where
\bb\mm_G:=\mm-\int_G
\rho_L(g^{-1})\mm\rho_R(g)\de g.\eee
 Note that $\mm-\mm_G$ leads to gauge
invariant mass terms and $G_{\mm_G}=G_\mm$.
We now introduce the change of variables
\bb\Phi:=  H-i\dd_G=:i\pmatrix {0&\varphi \cr
\varphi^*&0}\in\Omega_\dd^1\aa\label{fi}\ee
with $\varphi=h-\mm_G$. Then, assuming of course
gauge invariant internal Dirac operator, $\dd^g=\dd$,
$\Phi$ is homogeneously transformed into
\bb\Phi^g&=&H^g-i\dd^g_G
=\rho(g)[H-i\dd]\rho(g^{-1})+i\dd-i\dd+
i\int_G\rho(
g'^{-1})\dd\rho(g')\de g'\cr
&=&\rho(g)\,\left[H-\left(i\dd-i\int_G\rho(
g'^{-1})\dd\rho(g')\de
g'\right)\right]\,\rho(g^{-1})
=\rho(g)\Phi\rho(g^{-1}),\label{hom}\ee
and \bb\varphi^g =
\rho_L(g)\varphi\rho_R(g^{-1}).\eee
Now $h=0$, or
equivalently $\varphi=-\mm_G$, is certainly a
minimum of the preliminary Higgs potential and this
minimum spontaneously breaks $G$ if it is gauge
variant and non-degenerate.

Consider two extreme classes of examples,
vector-like and left-right models.

A {\it vector-like model} is defined by an arbitrary
internal algebra $\aa$ with identical left and right
representations, $\rho_L=\rho_R$, and with a mass
matrix proportional to the identity in each irreducible
component. As we shall see, every vector-like model
produces a Yang-Mills model with unbroken parity and
unbroken gauge symmetry, $G_\mm=G_\ell=G$,
 as electromagnetism or
chromodynamics. Since $\dd$ and $\rho$
commute, the internal differential algebra is trivial,
$\Omega_\dd^p\aa=0$ for $p\geq 1$, and the space of
Higgses is zero, $H=0$. The new variable $\Phi$
vanishes as well, because $\dd_G$ vanishes:
\bb\int_G\rho_L(g^{-1})\mm\rho_R(g)\de g
=\int_G\rho_L(g^{-1})\mm\rho_L(g)\de g
=\int_G\rho_L(g^{-1})\rho_L(g)\mm\de g
=\int_G\mm\de g=\mm.\eee
The preliminary Higgs potential vanishes identically,
but its minimum is non-degenerate. In this example, the
simpler variable $\Phi=H-i\dd$ would not be in a vector
space, because $\dd\not\in\Omega_\dd^1\aa$.

We define a {\it left-right model} by an internal
algebra consisting of a sum of  a ``left-handed'' and a
``right-handed'' algebra, $\aa=\aa_L\op\aa_R$ with the
left-handed algebra acting only on left-handed
fermions and similarly for right-handed
\bb
\rho_L(a_L,a_R)=\rho_L(a_L,0),\quad
\rho_R(a_L,a_R)=\rho_R(0,a_R),\qq
a_L\in\aa_L, \ a_R\in\aa_R.\eee
Now, any non-vanishing fermion mass
matrix breaks the gauge invariance, $G_\mm\not=G,
\ \mm\not=0$. At the same time, the internal Dirac
operator is a 1-form, $\dd=\dd_G\ \in\Omega_\dd^1\aa$,
because
\bb\int_G\rho_L(g^{-1})\mm\rho_R(g)\de g
&=&\int_{G_L\times
G_R}\rho_L(g_L^{-1},1)\mm\rho_R(1,g_R)\de g_L\de
g_R \cr\cr
&=&\left(\int_{G_L}\rho_L(g_L^{-1},1)\de g_L\right)
\mm\left(\int_{G_R}\rho_R(1,g_R)\de g_R\right)=0.\eee
In left-right models, we have $\Phi=H-i\dd$.
A more interesting, intermediate example will be
discussed in section 3.1 below.

 In the next step, the vectors $\psi_L,\ \psi_R$, and $H$
are promoted to genuine fields, i.e. rendered spacetime
dependent. As already in classical quantum mechanics,
this is achieved by tensorizing with functions. Let us
denote by $\ff$ the algebra of
(smooth, real or complex valued) functions over
spacetime $M$. Consider the algebra $\aa_t:=\ff\ot\aa$.
The group of unitaries of the tensor algebra $\aa_t$ is
the gauged version of the group of unitaries of the
internal algebra $\aa$, i.e. the group of functions
from spacetime into the group $G$. Consider the
representation $\rho_t:=\ul\cdot\ot\rho$ of the
tensor algebra on the tensor product
$\hh_t:=\sss\ot\hh,$
where $\sss$ is the Hilbert space of square integrable
spinors on which functions act by multiplication:
$ (\ul f\psi)(x):=f(x)\psi(x)$, $ f\in\ff,\
\psi\in\sss$.
The definition of the tensor product of Dirac operators,
\bb\dd_t:=\ddd\ot 1+\gamma_5\ot\dd\eee
comes from non-commutative geometry. We now
repeat the above construction for the infinite
dimensional algebra $\aa_t$ and its K-cycle. As
already stated, for $\aa=\cc,\ \hh=\cc,\ \mm=0$ the
differential algebra $\Omega_{\dd_t}\aa_t$ is
isomorphic to the de Rham algebra of differential
forms $\Omega (M,\cc)$. For general $\aa$, using the
notations of  \cite{sz}, an anti-Hermitian 1-form
$ H_t\in\Omega_{\dd_t}^1\aa_t,$
\bb H_t=A+H,\eee
 contains two pieces, an
anti-Hermitian Higgs {\it field} $
H\in\Omega^0(M,\Omega_\dd^1\aa)$ and a genuine
gauge field $ A\in\Omega^1(M,\rho(\gg))$
with values in the Lie algebra of the group of
unitaries,
 $ \gg:=\left\{ X\in\aa,\ X^*=-X\right\},$
represented on $\hh$. The curvature of $H_t$
\bb C_t:=\delta_tH_t+H_t^2\ \in\Omega_\dd^2\aa_t\eee
contains three pieces,
\bb C_t=C+F-\dee\Phi\gamma_5,\eee
 the ordinary, now $x$-dependent,
curvature $C=\delta H+H^2$, the field strength
\bb F:=\de A+\frac{1}{2}[A,A]\quad \in
\Omega^2(M,\rho(\gg))\eee
and the covariant derivative of $\Phi$
\bb\dee \Phi=\de \Phi+[A,\Phi]
\quad\in\Omega^1(M,\Omega_\dd^1\aa).\eee
Note that the covariant derivative may be applied to
$\Phi$ thanks to its homogeneous transformation law,
equation (\ref{hom}).

The definition of the Higgs potential in the
infinite dimensional space
\bb V_t(H_t):=(C_t,C_t)\eee
requires a suitable regularisation of the sum of
eigenvalues over the space of spinors $\sss$.
Here, we have to suppose spacetime to be compact and
Euclidean. Then the regularisation is
achieved by the Dixmier trace which allows an explicit
computation of $V_t$.
One of the miracles in CL models is that $V_t$ alone
reproduces the complete bosonic action of a YMH
model. Indeed it consists of three pieces,
the Yang-Mills action, the covariant Klein-Gordon
action and an integrated Higgs potential
\bb V_t(A+H)=\int_M\t (F*F\,z)+ \int_M\t
(\dee\Phi^**\dee\Phi\,z)+ \int_M*V(H).\label{Vt}\ee
As the preliminary Higgs potential $V_0$, the (final)
Higgs potential $V$ is calculated as a function of the
fermion masses,
   \bb V:=V_0-\t[\alpha C^*\alpha C\,z]=
\t[(C-\alpha C)^*(C-\alpha C)\,z].\eee
The linear map
$   \alpha: \Omega_\dd^2\aa\longrightarrow
         \rho(\aa)+\pi(\delta(\ker\pi)^1)$
is determined by the two equations
\bb
\t\lb R^*(C-\alpha C)\,z\rb&=&0\qquad{\rm for\ all}\
R\in\rho(\aa), \\
 \t\lb K^*\alpha C\,z\rb &=&0\qquad {\rm for\ all}\
K\in\pi(\delta(\ker\pi)^1).\eee
All remaining traces are over the finite dimensional
Hilbert space $\hh$. Note the ``wrong'' signs of
the first and third terms in equation (\ref{Vt}). The
signs are in fact correct for Euclidean spacetime.

Another miracle happens in the fermionic sector,
where the completely covariant action
 $\psi^*(\dd_t+iH_t)\psi$
reproduces the complete fermionic action of a YMH
model.
We denote by
 \bb\psi=\psi_L+\psi_R\ \in
\hh_t=\sss\,\ot\,\left(\hh_L\op\hh_R\right)\eee
 the multiplets of spinors and by $\psi^*$ the dual of
$\psi$ with respect to the scalar product of the
concerned Hilbert space. Then
\bb\psi^*(\dd_t+iH_t)\psi&=&
\int_M*\psi^*(\ddd+i\gamma(A))\psi
-\int_M*\left(\psi_L^*h\gamma_5\psi_R
+\psi_R^*h^*\gamma_5\psi_L\right)\cr\cr
&&\qq+\int_M*\left(\psi_L^*\mm\gamma_5\psi_R
+\psi_R^*\mm^*\gamma_5\psi_L\right)\cr\cr
&=&\int_M*\psi^*(\ddd+i\gamma(A))\psi
-\int_M*\left(\psi_L^*\varphi\gamma_5\psi_R
+\psi_R^*\varphi^*\gamma_5\psi_L\right)\cr\cr
&&\qq
+\int_M*\left(\psi_L^*(\mm-\mm_G)\gamma_5\psi_R
+\psi_R^*(\mm-\mm_G)^*\gamma_5\psi_L\right)
\label{diract}\ee
containing the ordinary Dirac action
and the Yukawa couplings.
If the minimum $\varphi=v$ is
non-degenerate, we retrieve the input fermionic mass
matrix $\mm$ on the output side by setting the
perturbative variables $h$ to zero in the first equation
in (\ref{diract}). The rhs of the second equation in
(\ref{diract}) is the fermionic action written with the
homogeneous scalar variable $\varphi$. The second
term yields the trilinear invariants (\ref{tri})
with Yukawa couplings fixed such that
$\mm$ is the fermionic mass matrix. As already pointed
out, the third term is an invariant mass term and
therefore admissible in a YMH Lagrangian.
Consequently every CL model with non-degenerate
vacuum is a YMH model with
$\hh_S=\left\{ H\in \Omega_\dd^1\aa,\ H^*=-H
\right\}.$
Note that $\hh_S$ carries a group representation, that
is not necessarily an algebra representation and
we have the following inclusion of group
representations
$
\hh_S\:\subset\:\left(\hh_L^*\ot\hh_R\right)\,\op\,
\left(\hh_R^*\ot\hh_L\right).$
Furthermore
$ G_\ell=G_\mm=G_{\mm_G}.$
We have nothing to
say about degenerate vacua i.e. minima of the Higgs
potential, that lie on distinct gauge orbits. In fact,
whether these are allowed in YMH models is a question
of taste for some, a question of quantum corrections for
others. We shall indicate a few examples. A final
remark concerns the unusual appearance of
$\gamma_5$ in the fermionic action (\ref{diract}).
Just as the ``wrong'' signs in the bosonic action
(\ref{Vt}), these $\gamma_5$ are proper to the
Euclidean signature and disappear in the Minkowski
signature.

Thomas Schucker
CPT, case 907
F-13288 Marseille
cedex 9
tel.: (33) 91 26 95 32
fax:  (33) 91 26 95 53

\section{Examples with degenerate vacua}

\subsection{Discrete degeneracy}

Our first example is in between
vector-like models, $\mm_G=0$, and left-right models,
$\mm_G=\mm$, in the sense that here $\mm_G\not=0$
and $\not=\mm$.  Choose as internal algebra
$\aa=M_2(\cc)$, the algebra of complex $2\times 2$
matrices. Both left- and right-handed fermions come in
$N$ generations of doublets,
$ \hh_L=\hh_R \:=\ \cc^2\otimes\cc^N$.
These Hilbert spaces carry identical left and right
representations
\bb
\rho_L(a)=\rho_R(a):=a\otimes 1_N,\quad a\in\aa.\eee
The fermion mass matrix is chosen block
diagonal to ensure conservation of the electric charge,
$G_\mm=U(1)$:
\bb \mm=\pmatrix{
m_1&0\cr
0&m_2},\eee
 $m_1$ and $m_2$ are complex $N\times N$
matrices which should be thought of as mass matrices
of the quarks of electric charge 2/3  and -1/3 and we
suppose them different, $m_1\neq m_2$. Then
\bb\mm_G&=&\mm-\int_{U(2)}(g^{-1}\ot 1_N)\mm
(g\ot1 _N) \de g
\cr\cr &=& 1_2\ot\frac{1}{2}(m_1+m_2)+\sigma_3\ot
\frac{1}{2}(m_1-m_2)\cr\cr&&-
\int_{U(2)}g^{-1}1_2 g \de g\ot\frac{1}{2}(m_1+m_2)
-\int_{U(2)}g^{-1}\sigma_3 g \de g\ot\frac{1}{2}
(m_1-m_2)
\cr\cr&=&
\frac{1}{2}\sigma_3\ot\mu\eee
where we have put
$\sigma_3:=\pp{
1&0\cr 0&-1},\
\mu:=m_1-m_2,$
and we have used the identity
\bb \int_{U(2)}g^{-1}A g \de g=\frac{1}{2}(\t A)\,1_2,
\qq A \in M_2(\cc).\eee
A general component of the Higgs field
takes the form $h=h_1\ot\mu$, $h_1$ being an
arbitrary Hermitian $2\times 2$-matrix. Likewise
$\varphi=\varphi_1\ot\mu$ and
$\varphi_1=h_1-1/2\sigma_3$. In these variables, the
Higgs potential can be computed
 to be \cite{is} \bb
V(H)=2\lp\t\lp(\mu\mu^*)^2\rp-\frac{(\t\mu\mu^*)
^2}{N}\rp\t\lb\left(\varphi_1+1_2/2\right)^2
\left(\varphi_1-1_2/2\right)^2\rb.\eee
$z$ is necessarily a positive scalar and we have put
$z=1_{4N}$. For $N=1$ generation, the Higgs potential
vanishes identically, and any point in $\hh_S$ is
minimum. The situation is more exciting in presence
of two or more generations . Then, the minima lie on
three disconnected pieces, the orbit of
$\varphi=-\mm_G$ with little group
$G_\ell=G_\mm=U(1)$, and two isolated points
$\varphi_1=\pm1_2/2$ with little group $G_\ell=U(2)$.
We may wonder if quantum corrections \cite{cw} do
lift this degeneracy and if so, in favor of which
vacuum.

Since $\dd_G$ is a 1-form, one can compute the
curvature of the Higgs $i\dd_G$:
\bb \delta\left(i\dd_G\right)\,+\,\left(i\dd_G\right)^2
=\frac{1}{2}\left(\t
\left[\mm\mm^*\right]-\frac{1}{2}|\t \mm|^2\right)
\,1_4\ot 1_N,\qq {\rm for\ all}\ N.\eee
Remark that, in the similar looking model by
M. Dubois-Violette, R. Kerner \& J.
Madore \cite{dvkm}, this curvature vanishes.

\subsection{Continuous degeneracy}

In the last example we had a finite, discrete
degeneracy: the vacuum consisted of three
disconnected orbits. Now we would like to present
a left-right model with continuous degeneracy, the
orbits of minimum will lie on a horizontal gutter.
Consider the complex algebra
$\aa=M_2(\cc)\op\cc\op\cc$ with representations on
$\hh_L=\cc^2$, $\hh_R=\cc^2$ given by
\bb\rho_L(a,b,b')=a, \qq
\rho_R(a,b,b')=\pp{b&0\cr 0&b'}=:B,\qq  a\in M_2(\cc),
\ (b,b')\in\cc\op\cc. \eee
Let the mass matrix be as in the last example with one
generation,
\bb \mm=\pp{m_1&0\cr 0&m_2},\qq
m_1,m_2\in\cc,\ |m_1|\not=|m_2|.\eee
Recall that for any left-right model we have
$\mm_G=\mm$ and $\dd_G=\dd$.
A general element of $\Omega_\dd^1\aa$ is of the
form
\bb
\pi((a_0,b_0,b'_0)\delta(a_1,b_1,b'_1)) &=& i\pmatrix
{0&a_0(a_1-B_1)\mm\cr
-\mm^*B_0(a_1-B_1)&0}\cr\cr &=&H=i\pp{0&h\cr
\tilde h^*&0}= \pmatrix
{0&h_1\mm\cr \mm^*\tilde h_1^*&0},\ h_1,\tilde
h_1\in M_2(\cc).\eee
As element of
$\pi(\Omega^2\aa)$, $\delta H$ is
\bb
\delta H&=&
\pi(\delta(a_0,b_0,b'_0)\delta(a_1,b_1,b'_1))\cr \cr
&=&\pmatrix {(a_0-B_0)\mm\mm^*(a_1-B_1)&0\cr
0&\mm^*(a_0-B_0)(a_1-B_1)\mm}\cr  \cr
&=&\pmatrix
{\Sigma (a_0-B_0)(a_1-B_1)+ \Delta
(a_0-B_0)\sigma_3(a_1-B_1)&0\cr
0&\mm^*(a_0-B_0)(a_1-B_1)\mm}\eee

 where we have
used the decomposition
\bb   \mm\mm^* = \pmatrix
{|m_1|^2&0\cr 0&|m_2|^2} = \Sigma
1_2+\Delta\sigma_3\eee   with
\bb   \Sigma := {1\over2}(|m_1|^2+|m_2|^2),\qq
   \Delta := {1\over2}(|m_1|^2-|m_2|^2).\eee
A general element in
$(\ker\pi)^1$ is a finite sum of the form
$\sum_j
(a^j_0,b^j_0,b'^j_0)\delta (a^j_1,b^j_1,b'^j_1)$
 with the two conditions
\bb   \left[ \sum_j a^j_0(a^j_1-B^j_1)\right]\mm =
0,\qq
   \mm^*\left[ \sum_j B^j_0(a^j_1-B^j_1)\right] =
0.\eee   Therefore the corresponding general element in
$\pi(\delta (\ker\pi)^1)$ is
\bb  \delta H= \pmatrix
{\Sigma\sum_j(a^j_0-B^j_0)(a^j_1-B^j_1)+
\Delta\sum_j(a^j_0-B^j_0)\sigma_3(a^j_1-B^j_1)&0\cr
0&0} \eee
 still subject to the two conditions.  Recall that
$\Delta \ne 0$ by assumption and we have  the
following inclusion
\bb   \pi(\delta(\ker\pi)^1) \supset
\left\{\pmatrix {\Delta\sum_j a^j_0\sigma_3
a^j_1&0\cr 0&0},\quad \sum_j a^j_0a^j_1=0 \right\}
= \left\{ \pmatrix {\Delta k&0\cr 0&0}, \quad k \in
M_2(\cc)\right\}.\eee
   To prove the last equality, we
note that the subspace is a two-sided  ideal in the rhs
and non-zero. The
algebra $M_2(\cc)$ being simple the subspace is the
whole algebra. Consequently the junk is
\bb  J^2= \pi(\delta(\ker\pi)^1)
= \left\{ \pmatrix { k&0\cr 0&0}, \quad k \in
M_2(\cc)\right\}.\eee
 Now  we compute the quotient
$ \Omega_\dd^2\aa = \pi(\Omega^2\aa)/J^2$
as orthogonal complement of the junk in
$\pi(\Omega^2\aa)$ with respect to the scalar
product (\ref{sp})
with $z=1_4$,
\bb   \Omega_\dd^2\aa = \left\{
\pmatrix {0&0\cr 0&\mm^*c_1\mm}, \quad c_1 \in
M_2(\cc)\right\}.\eee
Let us recapitulate:
\bb   \Omega_\dd^0\aa = \left\{
\pmatrix {a&0\cr 0&B}, \quad a \in M_2(\cc),\
B=\pmatrix {b&0\cr 0&b'}\right\},\eee
\bb   \Omega_\dd^1\aa = \left\{i\pmatrix {0&h_1\mm\cr
\mm^*\tilde h_1^*&0},\  h_1,\tilde h_1\in
M_2(\cc)\right\},\eee
\bb   \Omega_\dd^2\aa = \left\{
\pmatrix {0&0\cr 0&\mm^*c\mm}, \quad c \in
M_2(\cc)\right\}.\eee
Since $\pi$ is a
*-homomorphism, the product  in
$\Omega_\dd\aa$ is given by matrix multiplication
followed by the  projection
\bb P=\pmatrix{0&0\cr
0&1_2}\eee
 and the involution  is
given by transposition and complex conjugation. In
order to
calculate the  differential $\delta$, we went back to the
differential envelope:
\bb \delta:\pmatrix
{a&0\cr 0&B}\in\Omega_\dd^0\aa &\longmapsto&
 i\pmatrix
{0&(a-B)\mm\cr -\mm^*(a-B)&0}\in\Omega_\dd^1\aa,
\cr\cr \delta :
 i\pmatrix {0&h_1\mm\cr \mm^*\tilde
h_1^*&0}\in\Omega_\dd^1\aa &\longmapsto&
\pmatrix {0&0\cr
0&\mm^*(h_1+\tilde h_1^*)\mm}\in\Omega_\dd^2\aa.
\eee
Let now
\bb  H =
i\pp{0&h\cr h^*&0}=
i\pmatrix {0&h_1\mm\cr \mm^*h_1^*&0}, \quad
h_1\in M_2(\cc),\eee
be a Higgs. Its homogeneous variable is
\bb  \Phi:= H-i\dd_G= H-i\dd \
= i\pmatrix {0&\varphi\cr \varphi^*&0}
=i\pmatrix {0&\varphi_1\mm\cr
\mm^*\varphi_1^*&0} .\eee
 In other words
$\varphi_1=h_1-1_2$
is an arbitrary, complex $2\times 2$ matrix.
Under the group of unitaries $G=U(2)\times U(1)\times
U(1)$, it still decomposes into two irreducible
pieces, its two column vectors, $\varphi_1=:
(\varphi_{11},\varphi_{12})$.
In terms of these variables, the curvature reads
\bb   C := \delta H+H^2 =
\pmatrix {0&0\cr 0&\mm^*c\mm}\ \
\in\Omega_\dd^2\aa\eee
 with
$c = h_1+h_1^*-h_1^*h_1
= 1_2-\varphi_1^*\varphi_1.$
The preliminary Higgs potential is
\bb V_0(H)&=& \t\lb C^2\rb =
\  \t\lb\left(
\mm^*(1_2-\varphi_1^*\varphi_1)\mm\right)^2\rb
\cr &=&|m_1|^4+|m_2|^4+|m_1|^4(\varphi_{11}^*
\varphi_{11})^2+
|m_2|^4(\varphi_{12}^*\varphi_{12})^2 \cr
             &&\ -2|m_1|^4\varphi_{11}^*\varphi_{11}
                -2|m_2|^4\varphi_{12}^*\varphi_{12}
+2|m_1|^2|m_2|^2(\varphi_{11}^*\varphi_{12})
(\varphi_{12}^*\varphi_{11}).\eee
Its minimum is non-degenerate and  spontaneously
 breaks the gauge symmetry. However with
\bb   \alpha C=\pmatrix{ 0&0&0&0\cr 0&0&0&0\cr
0&0&|m_1|^2c_{11}&0\cr 0&0&0&|m_2|^2c_{22}},\eee
 the Higgs potential
\bb   V=\t\lb ( C-\alpha C)^2\rb=
2|m_1|^2|m_2|^2\left|\varphi_{11}^*\varphi_{12}
\right|^2\eee
  has  continuously degenerate vacua which also
include the gauge invariant point, $\varphi_{11}=
\varphi_{12}=0$. Indeed, the Higgs potential vanishes
if and only if the two complex doublets $\varphi_{11}$
and $ \varphi_{12}$ are orthogonal, irrespective of
their lengths. Finally, we remark that the Higgs
potential has only symmetry breaking minima for two
and more generations.

\subsection{Complete symmetry breakdown}

We have seen that, in CL models with non-degenerate
vacuum, the little group coincides with $G_\mm$. The
latter is controlled immediately by the input. We take
advantage of this to construct a model with complete,
spontaneous symmetry breakdown, i.e. finite little
group. Consider a left-right model with {\it real}
internal algebra $\aa=\hhh\op\cc$, $\aa_L=\hhh$
being the
quaternions, and two generations of fermions
\bb \hh_L=\cc^2\,\ot\,\cc^2,&\quad\rho_L(a,b)=a\ot
1_2 =\pp{a&0\cr0&a},&
\ a\in\hhh,\cr \cr
\hh_R=(\cc\op\cc)\,\ot\,\cc^2,&\quad
\rho_R(a,b)=B\ot 1_2=\pp{B &0\cr0&B},&\ b\in\cc,\
B:=\pp{b&0\cr 0&b^*}. \eee
We choose the mass matrix
\bb\mm=\pp{
\mm_1&0\cr
0&\mm_2},\qq
\mm_1:=\pp{m_1&0\cr0&m_2}, \quad
\mm_2:=\pp{0&m\cr m&0},\eee
with
$m_1,m_2,m\in\rr,\ m_1\not=
m_2, \ m\not=0.$
Therefore $G_\mm=\zz_2$. A general 1-form
is a finite sum of terms
\bb H= -i\pi((a_0,b_0)\delta(a_1,b_1))=i\pp{
0&0&h_1\mm_1&0\cr
0&0&0&h_2\mm_2\cr
\mm_1\tilde h_1&0&0&0\cr
0&\mm_2\tilde h_2&0&0}\eee
with
\bb & h_1:=a_0(a_1-B_1),\quad
&h_2:=a_0(a_1-B^*_1),\cr
&\tilde h_1:=-B_0(a_1-B_1),
\quad &\tilde h_2:=-B_0^*(a_1-B_1^*),\eee
After the finite summation, the four quaternions
$h_1,h_2,\tilde h_1,\tilde h_2$ are independent
in general.
 The junk in degree two is
\bb \pi(\delta(\ker\pi)^1)=\left\{\pp{
i(m_1^2-m_2^2)k&0&0&0\cr
0&0&0&0\cr
0&0&0&0\cr
0&0&0&0},\ k\in\hhh\right\}\eee
and
\bb \delta H=\pp{
\frac{1}{2}(m_1^2+m_2^2)(h_1+\tilde h_1)&0&0&0\cr
0&m^2(h_2+\tilde h_2)&0&0\cr
0&0&\mm_1(h_1+\tilde h_1)\mm_1&0\cr
0&0&0&\mm_2(h_2+\tilde h_2)\mm_2}.\eee
A Higgs, an anti-Hermitian 1-form, is characterized by
two independent quaternions, $h_1$ and $h_2$,
\bb h=\pp{h_1\mm_1&0\cr 0&h_2\mm_2},\quad
\varphi=\pp{\varphi_1\mm_1&0\cr
0&\varphi_2\mm_2},\eee
 with $\varphi_j=h_j-1_2,\ j=1,2$. Let us
decompose each quaternion
\bb\varphi_j=\pp{x_j&-y_j^*\cr y_j&x_j^*},\quad
x_j,y_j\in\cc \eee
into its two column vectors
\bb\varphi_j=\left(\varphi_{j1},\
-i\sigma_2\varphi^{*T}_{j1}\right),\quad
\varphi_{j1}= \pp{x_j\cr y_j},\quad
\sigma_2:=\pp{0&-i\cr i&0}. \eee
They define the irreducible pieces of the Higgs
under a unitary transformation $g=(g_2,g_1)\in SU(2)
\times U(1)$:
\bb\varphi_{11}^g=g_2\varphi_{11}g_1^{-1},\quad
\varphi_{21}^g=g_2\varphi_{21}g_1.\eee
In other words the Higgs consists of two complex
$SU(2)$-doublets with opposite $U(1)$-charges. Note
that if $\mm_2$ was also diagonal, we would only have
one complex Higgs
doublet. Now the computation of the Higgs potential is
lengthy, but straightforward. In terms of the two
doublets, the result for $z=1_8$ is
\bb V(H)=(m_1^4+m_2^4)
\left[1-\varphi_{11}^*\varphi_{11}\right]^2
+2m^4\left[1-\varphi_{21}^*\varphi_{21}\right]^2
-(m_1^2+m_2^2)m^2
\left[1-\varphi_{11}^*\varphi_{11}\right]
\left[1-\varphi_{21}^*\varphi_{21}\right].\eee
The Higgs potential is zero if and only if both complex
doublets $\varphi_{11}$ and $\varphi_{21}$ have
length one. Since their relative orientation is
arbitrary and gauge invariant, the vacua are
continuously degenerate. However, in  every vacuum,
all four gauge bosons are massive and the four masses
are independent of the relative orientation.
Furthermore, the little groups of all vacua are equal,
$G_\ell=\left\{-1,+1\right\}$, as expected.

\section{ Necessary conditions}

One may very well do general relativity using only
Euclidean geometry. Still, we agree that
Riemannian geometry is the natural setting of general
relativity. A main argument in favor of this attitude
is that there are infinitely more gravitational theories
in Euclidean geometry than in Riemannian geometry.
The same is true for the standard model. Its natural
setting, to our taste, is non-commutative geometry.
The fact that today's Yang-Mills-Higgs model of
electro-weak and strong interactions falls in the
infinitely smaller class of Connes-Lott models is
remarkable. The purpose of this section is to show in
what extent it is remarkable. We give a list of
constraints on the input of a YMH model. They are
necessary conditions for the existence of a
corresponding CL model.

\subsection{The group}

The compact Lie group $G$ defining a Yang-Mills
model must be chosen such that its Lie algebra $\gg$
admits an invariant scalar product. Therefore $\gg$ is a
direct sum of simple and abelian algebras. After
complexification, the simple Lie algebras are classified
according to E. Cartan, into four infinite series,
$su(n+1),\ n\geq 1,\qq o(2n+1),\ n\geq 2,\qq sp(n),\
n\geq 3,\qq o(2n),\ n\geq 4$ and five exceptional
algebras $G_2,\ F_4,\ E_6,\ E_7,\ E_8$. To define a CL
model, we need a real or complex involution algebra
$\aa$ admitting a finite dimensional, faithful
representation. Their classification is easy. In the
complex case, such an algebra is a direct sum of matrix
algebras $M_n(\cc),\  n\geq 1$. In the real case, we
have direct sums of matrix algebras with real, complex
or quaternionic coefficients, $M_n(\rr),\ M_n(\cc),\
M_n(\hhh),\qq n\geq 1$. The corresponding groups
of unitaries are $O(n,\rr),\ U(n),\ USp(n)$. Note the
two isomorphisms, $USp(2)\cong SU(2)$ and
$USp(4)/\zz_2\cong SO(5,\rr)$.

Let us outline the proof of the classification. Since
$\aa$ has a faithful representation on a Hilbert space it
is semi-simple \cite{r}. Then $\aa$ is a finite sum of
$n\times n$ matrices over finite dimensional division
algebras \cite{j1}. There are only three finite, real
division algebras, $\rr,\ \cc$ and $\hhh$ \cite{j2}.

The groups accessible in a CL model therefore belong
to the second, third, and forth Cartan series.
Furthermore we have $u(n)\cong su(n)\op u(1)$. Up
to the $u(1)$ factor, this is the first series. At the group
level, this factor is disposed of by a condition on the
determinant. In the algebraic setting there is a similar
condition, that reduces the group of unitaries to a
subgroup, here $SU(n)$. This condition is called
unimodularity and is discussed in the next section. To
sum up, all classical Lie groups are accessible in a CL
model but the exceptional ones.

\subsection{The fermion representation}

In a YMH model, the left- and right-handed fermions
come in unitary representations of the chosen group
$G$. Every $G$ has an infinite number of irreducible,
unitary representations. They are classified by their
maximal weight. On the other hand, the above
involution algebras $\aa$ admit only one or two
irreducible representations. The reason is that an
algebra representation has to respect the multiplication
and the linear structure, while a group representation
has to respect only the multiplication. In particular,
the tensor product of two group representations is a
group representation, while the tensor product of two
algebra representations is not an algebra
representation, in general.

 The only irreducible representation of
$M_n(\cc)$ as complex algebra is the fundamental one
on $\hh=\cc^n$. Also $M_n(\rr)$ and $M_n(\hhh)$
have only the fundamental representations on
$\hh=\rr^n$ and $\hh=\cc^n\ot\cc^2$ while
$M_n(\cc)$ considered  as real algebra has two
inequivalent, irreducible representations, the
fundamental one: $\hh=\cc^n$, $\rho_1(a)=a$, $a\in
M_n(\cc)$, and its conjugate: $\hh=\cc^n$,
$\rho_2(a)=\bar a$.

The proof of this classification relies on the facts
that the centers of the above algebras $\aa=M_n(\rr),\
M_n(\cc),\ M_n(\hhh)$ are the corresponding division
algebras, $\rr,\ \cc,\ \rr$, and that the
representations of $\aa$ are classified (up to
equivalence) by the automorphisms of their centers
(as real algebras). Thus $M_n(\cc)$ is the only case
with two inequivalent representations
(Skolem-Noether theorem \cite{j3}).

Let us summarize. The only possible representations for
fermions in a CL model are
\begin{itemize}
\item
for $G=O(n,\rr)$, $N$ generations of the
fundamental representation on $\hh=\rr^n\ot\rr^N$,
 \item
for $G=U(n)$ (or $SU(n)$ ), $N$ generations of the
fundamental representation on $\hh=\cc^n\ot\cc^N$
and $\bar N$ generations of its conjugate on
$\hh=\cc^n\ot\cc^{\bar N}$.
\item
for $G=USp(n)$, $N$ generations of the
fundamental representation on
$\hh=\cc^n\ot\cc^2\ot\cc^N$,
\end{itemize}

In a YMH model with $G=SU(2)$, the fermions can be
put in any irreducible representations of dimension 1,
2, 3,... while in the corresponding CL model with
$\aa=\hhh$, there is only one irreducible
representation accessible for the fermions, the
fundamental, two dimensional one. Similarly, in a YMH
model with $G=U(1)$ the fermions can have any
(electric) charge from $\zz$ or even from $\rr$ if we
allow `spinor' representations. In the corresponding
CL model with $\aa=\cc$, fermions can only have
charges plus and minus one. In any case, if we want
more fermions in a CL model, we are forced to introduce
families of fermions.

\subsection{The gauge coupling constants}

In a YMH model, the gauge coupling constants
parameterize the most general gauge invariant scalar
product on the Lie algebra $\gg$ of $G$. In a CL model,
see the rhs of equation (\ref{Vt}), this
scalar product is not general but comes from the trace
over the fermion representation space $\hh$, equation
(\ref{gsp}). The scalar product involves the positive
operator $z$, that commutes with the internal Dirac
operator and with the fermion transformations
$\rho(\aa)$ and that leaves $\hh_L$ and $\hh_R$
invariant. Depending on the details of the mass matrix
and of the left- and  right-handed
 representations $\rho_L$ and $\rho_R$, the
gauge coupling constants may be constraint or not. The
examples of the last section will illustrate this point.

\subsection{ The Higgs sector}

As explained in section 2, the scalar representation
$\rho_S$ on $\hh_S$ in a CL model is a representation
of the {\it group} of unitaries only. This
representation is not chosen but it is calculated as a
function of the left- and right-handed fermion
representations and of the mass matrix. As illustrated
by the examples of section 3, the dependence of the
scalar representation on this input is involved
and we can make only one general  statement:
\bb\hh_S\:\subset\:\left(\hh_L^*\ot\hh_R\right)\,\op\,
\left(\hh_R^*\ot\hh_L\right).\eee
Nevertheless, this inclusion is sufficient to rule out the
possibility of spontaneous parity breaking in left-right
symmetric models \`a la Connes-Lott \cite{is}.

The Higgs potential as well, is on the output side of a CL
model. Its calculation involves the positive
operator $z$ from the input and is by far, the most
complicated calculation in this scheme. We know that
$\varphi=-\mm_G$ is an absolute minimum of the
Higgs potential. If it is non-degenerate, the gauge and
scalar boson masses are determined by the fermion
masses and the entries of $z$. See the last section for
examples.

Our last necessary condition concerns the Yukawa
couplings. In a CL model, they are determined such that
$\mm$ is the fermionic mass matrix after spontaneous
symmetry breaking. Up to the $z$ dependent scalar
normalization in the bosonic action (\ref{Vt}), the
Yukawa couplings are all one. Normalization details
are relegated to the appendix.

\section{The unimodularity condition}

The purpose of the unimodularity condition is to
reduce the group of unitaries $U(n)$ to its subgroup
$SU(n)$. At the group level, this is easily achieved by
the condition $\det g=1$. However the determinant
being a non-linear function is not available at the
algebra level. We are lead to use the trace instead,
together with the formula
\bb \det e^{2\pi iX}\,=\, e^{2\pi i\t X}.\eee
Even in the infinite dimensional case, the connected
component $G^0$ of the unit in the group of unitaries
$G$ is generated by elements $g=e^{2\pi iX}$,
$X=X^*\in\aa$. The desired reduction is achieved by
using the phase, defined by \cite{hs}
\bb {\rm phase}_\tau (g):=\frac{1}{2\pi
i}\int_0^1\tau \left(g(t)\frac{\de\ }{\de t}
g(t)^{-1}\right)\,\de t,\eee
 where $\tau$ is a linear
form on $\aa$ satisfying
\bb\tau(1)\in\zz,\qq
\tau(a^*)=\tau(a)^*,\qq \tau(a)=\tau(g^*ag),\qq g\in
G,\  a\in\aa^+:=\{bb^*,\ b\in\aa\},\eee
 and where $g(t)$ is a curve in $G^0$
connecting the unit to $g$. We obtain the finite
dimensional case above by putting
$ \tau(a)=\t \rho(a)$ and $g(t)=e^{2\pi
iXt}$.
The definition of the phase involves two
choices, that are easily controlled in finite dimensions:
the most general linear form $\tau$ can be written as
$\tau(a)=\t \rho (ap),\ a\in\aa,\ p\in{\rm
center}\, \aa,$
and the ambiguity in the choice of the curve $g(t)$ is
controlled by the first fundamental group $\pi^1(G^0)$
which is contained in $\zz$, see table below. Therefore
the unimodularity condition
\bb e^{2\pi i\,{\rm phase}_\tau(g)}=1\eee
is well defined and defines a subgroup
\bb G_p:=\left\{g\in G^0,\
e^{2\pi i\,{\rm phase}_{\t \rho (\cdot
p)}(g)}=1\right\}\eee
 of $G^0$. For $\aa=M_n(\cc)$,
$n\geq 2$, the center is spanned by $1_n$ and
$G_1=SU(n)$. The center of $\aa=M_n(\cc)\op
M_m(\cc)$, $n,m\geq 2$, is spanned by two elements,
$p_n$ and $p_m$,  the projectors on $M_n(\cc)$ and on
$M_m(\cc)$. We have
\bb G_{p_n}&=&SU(n)\times U(m),\cr
       G_{p_m}&=&U(n)\times SU(m),\cr
       G_{p_n+p_m}&=&S(U(n)\times U(m)).\eee
We close this section with a remark: the
described reduction of the group of unitaries $G$ to a
subgroup $G_p$ is compatible with the model building
kit of section 2. In particular
\bb \dd_{G_p}=\dd_{G^0}=\dd_G \label{ddgg}\ee
and the change of variables, equation (\ref{fi}), is
untouched. The proof of equations (\ref{ddgg}) is done
case by case and is summarized in the following table.
\bb\cr  \matrix{
G&G^0&G/G^0&G_1&G^0/G^1&\pi^1(G^0)\cr \cr
O(n,\rr)&SO(n,\rr)&\left\{{\rm
diag}(-1,1,...,1),1_n\right\}&SO(n,\rr)&\left\{1_n
\right\}&\zz_2\cr \cr
U(n)&U(n)&\left\{1_n\right\}&SU(n)&
\left\{e^{2\pi i/n}1_n,e^{4\pi i/n}1_n,...,
e^{n2\pi i/n}1_n\right\}&
\left\{1\right\}\cr \cr
USp(n)&USp(n)&\left\{1_{2n}\right\}&USp(n)&
\left\{1_{2n}\right\}&\left\{1\right\} }\eee
\vskip 1truecm\noindent
All elements of $G/G^0$ and $G^0/G_1$ are multiples of
the identity except for $O(n,\rr)/SO(n,\rr)$. However,
integrating $\rho(g^{-1})\dd\rho(g)$ first over the
normal subgroup $SO(n,\rr)$ yields a matrix whose
blocks are  already diagonal matrices.

Thomas Schucker
CPT, case 907
F-13288 Marseille
cedex 9
tel.: (33) 91 26 95 32
fax:  (33) 91 26 95 53

\section{The standard model}

We would like to conclude by locating the standard
model within the CL scheme. {\it The} pedagogical
example to illustrate the YMH model building kit is the
Georgi-Glashow $SO(3)$ model \cite{gg}. Miraculously
enough, {\it the} pedagogical example in the CL subkit
is almost the Glashow-Salam-Weinberg model. Indeed,
this example is the electro-weak algebra
$\aa=\hhh\op\cc$, (group of unitaries $G=SU(2)\times
U(1)$ ) represented on {\it two} generations of
leptons, $N=2$,
\bb \hh_L=\cc^2\ot\cc^2,\qq\hh_R=\cc\ot\cc^2.\eee
With respect to the suggestive basis
\bb \pp{\nu_e\cr e}_L,\ \pp{\nu_\mu\cr \mu}_L,\qq
e_R,\ \mu_R \eee
 of $\hh_L\op\hh_R$, the representation has the
following matrix form,
\bb \rho(a,b)=\pp{a\ot 1_N&0\cr 0&\bar b 1_N},\qq
a\in\hhh,\ b\in\cc.\eee
The internal Dirac operator is
\bb \dd=\pp{0&\mm\cr \mm^*&0}=\pp{
0&\pp{0\cr M_e}\cr
\pp{0&M_e}&0},\eee
with
\bb M_e:=\pp{m_e&0\cr 0&m_\mu}, \qq
m_e<m_\mu.\eee
The most general positive $6\times 6$ matrix $z$, that
commutes with $\rho(\aa)$ and with $\dd$ is
\bb z=\pp{1_2\ot\pp{y_1&0\cr 0&y_2} &0\cr
0&\pp{y_1&0\cr 0&y_2}}\eee
with positive numbers $y_1$ and $y_2$. Consequently
the coupling constants $g_2$ of $SU(2)$ and $g_1$ of
$U(1)$ are related,
\bb
\cot^2\theta_w=\left(\frac{g_2}{g_1}\right)^2
=2, \qq \sin^2\theta_w=1/3. \eee
Details are
given in the appendix. In this model, $\Phi$ of
equation (\ref{fi}) takes the form
\bb\Phi=i\pp{
0&\left[\pp{\varphi_1&-\bar\varphi_2\cr
\varphi_2&\bar\varphi_1}\ot\,1_N\right]\mm\cr
...^*&0}\label{Phi}\ee
and is parameterized by two functions $ \varphi_1,
\varphi_2:M\rightarrow\cc$. Under gauge
transformations, these transform as an $SU(2)$ doublet
\bb\varphi=\pp{\varphi_1\cr \varphi_2}.\eee
In terms of these parameters, the Higgs potential reads
\bb &&V(\varphi)= K\left(1-|\varphi|^2\right)^2,\cr
 K &:=&\frac{3}{2}
\left(y_1m^4_e+y_2m^4_\mu\right)
-\frac{3}{2} \frac{L^2}{y_1+y_2},\cr
 L&:=&y_1m^2_e+y_2m^2_\mu.\eee
Note that the scalar fields $\varphi_1$ and $\varphi_2$
are not properly normalized, they are dimensionless.
To get their normalization straight, we compute
the factor in front of the kinetic term $\t
\left(\de\Phi^**\de\Phi\, z\right)$
 in the Klein-Gordon action $\t \left(D\Phi^**D\Phi\,
z\right)$ as function of the variable $\varphi$. By
inserting equation (\ref{Phi}) we obtain:
\bb \t \left(\de\Phi^**\de\Phi\,
z\right)=*2L\left|\partial \varphi\right|^2.\eee
Likewise, we need the normalization of the gauge
bosons and as
shown in the appendix, we end up with the following
mass relations:
\bb m_W^2&=&\frac{L}{y_1+y_2}
=\frac{y_1m^2_e+y_2m^2_\mu}{y_1+y_2},
\cr  \cr
 m_H^2&=&\frac{2K}{L}
=3m^2_\mu\ \frac{
y_2}{y_1}\frac{(1-m_e^2/m_\mu^2)^2}
{(y_2/y_1+m_e^2/m_\mu^2)(y_2/y_1+1)}.\eee
Consequently \bb &&m_e\,<\,m_W\,<\,m_\mu,\cr
&& m_H\,<\,\sqrt{3}\left(m_\mu-m_e\right).\eee

We obtain a model with less constrained weak angle
by slightly modifying this example. Let us represent
the electro-weak algebra on one generation of leptons
and one generation of (uncoloured) quarks,
\bb\hh_L=\cc^2\ot\cc^2,\qq
\hh_R=\left(\cc\op\cc\right)\op\cc\eee
with suggestive basis
\bb\pp{u\cr d}_L,\qq
\pp{\nu_e\cr e}_L,\qq\qq
 \matrix{u_R,\cr d_R,}\qq e_R,\eee
and representation
\bb\rho(a,b):=\pp{
a&0&0&0\cr
0&a&0&0\cr
0&0&B&0\cr
0&0&0&\bar b},\qq(a,b)\in\hhh\op\cc,\qq
 B:=\pp{b&0\cr 0&\bar b}.\eee
We choose the internal
Dirac operator:
\bb\dd:=\pp{
0&0&\pp{m_u&0\cr 0&m_d}&0\cr
0&0&0&\pp{0\cr m_e}\cr
\pp{m_u&0\cr 0&m_d}&0&0&0\cr
0&\pp{0&m_e}&0&0}.\eee
All indicated fermion masses are supposed positive and
different. Now, the most general scalar product on the
differential algebra $\Omega_\dd(\hhh\op\cc)$ is
defined with the $7\times 7$ matrix
\bb z=\pp{x1_2&0&0&0\cr
0&y1_2&0&0\cr
0&0&x1_2&0\cr
0&0&0&y}\eee
with positive numbers $x$ and $y$. In this example we
 get:
\bb\sin^2\theta_w=\frac{x+y}{3x+2y},\eee
implying
\bb\frac{1}{3}\,<\,\sin^2\theta_w\,<\,\frac{1}{2}. \eee
 This $z$ is in the
image of the center of $\hhh\op\cc$ under $\rho$ if
and only if $x=y$ and we have
$\sin^2\theta_w=0.4$.

The drawback of these two examples --- electrically
charged neutrinos and up- and down-quark with
opposite charges --- is corrected by adding strong
interactions. As strong interactions are vector-like, this
addition is immediate except for the fact that the
representation of the left-handed quarks,
$(3,2,\frac{1}{3})$ in equation (\ref{HL}), is a tensor
product. However, this is a tensor product of two
representations of {\it two unrelated} algebras (
$M_3(\cc)$ and $\hhh$ ) and as such, it can be included
in the CL scheme by generalizing the representations
to bimodules \cite{cl,kB}. A bimodule is a pair of
algebras, each represented on a common Hilbert space,
such that the two representations commute. The
constraints indicated in section 4 remain otherwise
unaffected and for the standard model, they can be
stated as follows. The scalar representation is one weak
isospin doublet, implying a mass ratio for the $W$ and
$Z$ bosons given by the $\rho$ factor
\bb\rho:=\frac{m^2_W}{m^2_Z\cos^2\theta_w}=1.\eee
With the
general scalar product (\ref{gsp}), the other
constraints read \cite{ks2},
\bb m_t\,>\,\sqrt{3}\, m_W\,>\,\sqrt{3}\,m_e, \eee
\bb m_H= \sqrt{3\frac{(m_t/m_W)^4+2(m_t/m_W)^2-1}
{(m_t/m_W)^2+3}}\ m_W,\eee
 \bb\sin^2\theta_w&<&\frac{2}{3}\,\frac{m_t^2}
{m_t^2+m_W^2}.\eee
For the more restricted scalar product coming from the
center, the constraints are tighter:
\bb m_t&=&2\,m_W,\cr
m_H&=&3.14\,m_W,\cr
\sin^2\theta_w&<& \frac{8}{15}=0.533.\eee

\vskip1truecm
It is a pleasure to acknowledge help and advice of
Robert Coquereaux, Vaughan Jones, Daniel Kastler,
John Madore and Stanislaw Woronowicz.

\section{Appendix}

This appendix collects our normalization
conventions of a YMH model in a spacetime of
signature $+---$. Let $\varphi,\ \psi$, and
$W$ be complex fields of spin 0, 1/2, and 1. The kinetic
terms determine the normalization of the fields in the
Lagrangian and the masses and coupling constants are
defined with respect to this normalization. With
$\hbar=c=1$, the Lagrangian is
 \bb\cal{L}&=&
\frac{1}{2}\pa_\mu\varphi^*\pa^\mu\varphi
-\frac{1}{2}m^2_\varphi\varphi^*\varphi
+\bar\psi i\ddd\psi-m_\psi\bar\psi\psi\cr
&&-\frac{1}{2}\pa_\mu W^*_\nu\pa^\mu W^\nu
+\frac{1}{2}\pa_\mu W^{*\mu}\pa_\nu W^\nu
+\frac{1}{2}m^2_WW^*_\mu W^\mu.\eee
Note the one half in front of the scalar Lagrangian, i.e.
we decompose the complex scalar into real fields as
$\varphi={\rm Re}\varphi+i{\rm Im}\varphi$. We
use the following definitions:
\bb   \psi = \pmatrix{
\psi_1(x) \cr \psi_2(x) \cr \psi_3(x) \cr \psi_4(x)
},\qq \ddd\psi := \gamma^\mu
\pa_\mu\psi,\qq
\bar\psi:=(\psi_1^*,\psi_2^*,\psi_3^*,\psi_4^*) \,
\gamma^0.\eee
 Our gamma matrices are ,
\bb   \gamma ^0 :=  \pmatrix {1
& 0 & 0 & 0 \cr
 0 & 1 & 0 & 0 \cr
 0 & 0 & -1 & 0 \cr
 0 & 0 & 0 & -1} \qquad \gamma ^1 :=  \pmatrix {0 & 0 &
0 & 1 \cr
 0 & 0 & 1 & 0 \cr
 0 & -1 & 0 & 0 \cr
 -1 & 0 & 0 & 0}\eee
 \bb   \gamma ^2 :=  \pmatrix {0 & 0 & 0 &
-i \cr
 0 & 0 & i & 0 \cr
 0 & i & 0 & 0 \cr
 -i & 0 & 0 & 0} \qquad \gamma ^3 :=  \pmatrix {0 & 0 &
1 & 0 \cr
 0 & 0 & 0 & -1 \cr
 -1 & 0 & 0 & 0 \cr
 0 & 1 & 0 & 0}.\eee
They satisfy the anticommutation
relation
$\gamma ^\mu \gamma ^\nu +\gamma ^\nu
\gamma ^\mu =  2\eta ^{\mu \nu }1_4$
 with the flat
Minkowski metric
\bb   \eta = \pmatrix {1 & 0 & 0 & 0 \cr
 0 & -1 & 0 & 0 \cr
 0 & 0 & -1 & 0 \cr
 0 & 0 & 0 &-1 }.\eee
We take
\bb    \gamma_5 :=i
\gamma^0\gamma^1\gamma^2\gamma^3 =
 \pmatrix {0 & 0 & 1 & 0 \cr
 0 & 0 & 0 & 1 \cr
 1 & 0 & 0 & 0 \cr
 0 & 1 & 0 & 0}\eee
such that $\gamma_5^2=1_4$. $\gamma_5$
anticommutes with all other gamma matrices,
$\gamma^\mu\gamma_5+\gamma_5\gamma^\mu =
0$.
With the definitions
\bb \psi_L:=\frac{1_4-\gamma_5}{2}\,\psi,\qq
\psi_R:=\frac{1_4+\gamma_5}{2}\,\psi\eee
the free Dirac Lagrangian reads
\bb \ll_\psi=\overline{\psi_L}i\ddd\psi_L+
\overline{\psi_R}i\ddd\psi_R
-m_\psi\overline{\psi_L}\psi_R-
m_\psi\overline{\psi_R}\psi_L.\eee
In Euclidean spacetime, the Dirac Lagrangian written
in this chiral form vanishes identically and the
fermions have to be doubled. With
\bb W_{\mu\nu}:=\pa_\mu W_\nu-\pa_\nu W_\mu,
\eee
the free part of the Yang-Mills Lagrangian becomes
\bb\ll_W=-\frac{1}{4}W^*_{\mu\nu}W^{\mu\nu}
+\frac{1}{2}m_W^2 W^*_\mu W^{\mu}.\eee
The couplings of the gauge bosons to scalars and
fermions in their respective representations are
introduced through the covariant derivatives, while
the self couplings of the gauge bosons come from the
field strength.
All their coupling constants derive from the choice of
one invariant scalar product on the Lie algebra.
Amazingly enough, the parametrisation of this scalar
product seems uniform in the literature, at least for the
classical groups,
\bb (b,b')&:=&\frac{1}{g_1^2}\bar bb',
\qq b,b'\in\ u(1),
\cr\cr  (a,a')&:=&\frac{2}{g_n^2}\t(a^*a'),\qq
a,a'\in su(n). \eee
The gauge bosons sit in a 1-form $A=A_\mu\de
x^\mu$ with values in the Lie algebra and the
Yang-Mills Lagrangian reads
\bb\ll_{YM}= -\frac{1}{4}\left(F_{\mu\nu},
F^{\mu\nu}\right)\eee
with the field strength $F=1/2\,F_{\mu\nu}\de x^\mu
\de x^\nu,\
 F_{\mu\nu}=\pa_\mu A_\nu-\pa_\nu A_\mu+
\left[A_\mu,A_\nu\right].$

As an illustration, let us consider the standard model of
electro-weak interactions with $G=SU(2)\times U(1)$,
one doublet of scalars $\varphi$ and Higgs potential
 \bb V(\varphi) =
\lambda(\varphi^*\varphi)^2-\frac{\mu^2}{2}
(\varphi^*\varphi)\label{Va}.\ee
 First we choose the electric charge
generator $Q$:
\bb iQ:=i\left(g_2\sin\theta_w\pp{1/2&0\cr 0&-1/2}
,g_1\cos\theta_w\right),\eee
a normalized vector in the Cartan subalgebra of
$\gg=su(2)\op u(1)$ spanned by the weak isospin and
hypercharge,
\bb I_3:=i\left(g_2\pp{1/2&0\cr 0&-1/2},0\right),\qq
Y:=i\left(0,g_1\right).\eee
We complete $iQ$ to an orthonormal basis of
$\gg^\cc$ of eigenvectors of $[Q,\cdot]$
\bb \tilde Z&:=&
i\left(g_2\cos\theta_w\pp{1/2&0\cr 0&-1/2}
,-g_1\sin\theta_w\right),\cr \cr
I^+&:=&i\left(\frac{g_2}{\sqrt 2}\pp{0&1\cr
0&0},0\right),\cr \cr
 I^-&:=&i\left(\frac{g_2}{\sqrt
2}\pp{0&0\cr 1&0},0\right).\eee
The eigenvalues are 0 and $\pm
g_2\sin\theta_w=:\pm e$. The multiplet of gauge
bosons is now written as
 \bb A_\mu(x):= \gamma_\mu(x)\,iQ
+Z_\mu(x)\,\tilde
Z+\frac{1}{\sqrt{2}}\left(W_\mu(x)\,I^+
+W^*_\mu(x)\,I^-\right),\eee
where the photon $\gamma_\mu(x)$ and the
$Z_\mu(x)$ are real fields while the $W$
is complex.

The scalar fields sit in a $SU(2)$ doublet with
hypercharge $y_S=-1/2$:
\bb \tilde\rho_S(a,b)\varphi=
\left(a+y_Sb1_2\right)\varphi,\qq a\in su(2),
b\in u(1).\eee
$\tilde\rho_S$ denotes the Lie algebra representation.
In order to keep the photon massless, we must choose
$g_1$ such that one of the scalars has zero electric
charge,
\bb \frac{1}{i}\tilde\rho_S(iQ)=\pp{0&0\cr 0&-e}.\eee
This implies
\bb\frac{g_1}{g_2}=
\frac{\sin\theta_w}{\cos\theta_w}.\eee
The gauge bosons masses come from the absolute value
squared of the covariant derivative of the vacuum $v$.
Since $v$ satisfies $|v|^2=\mu^2/(4\lambda)$ we
choose \bb
v=\pp{\frac{1}{2}\sqrt{\frac{\mu^2}
{\lambda}}\cr 0}\eee and obtain
 \bb
\frac{1}{2}\left|\tilde\rho_S(A_\mu)v\right|^2=
\frac{1}{2}m_Z^2Z_\mu Z^\mu +\frac{1}{2}m_W^2
W^*_\mu W^{\mu} \eee
 with
\bb m_W=g_2\frac{\mu}{4\sqrt\lambda}\qq{\rm and}
\qq m_W=cos\theta_w m_Z.\eee
To compute the mass of the physical, real Higgs scalar
$H$, we change variables in the Higgs potential,
\bb \varphi= v+\pp{H(x)+ih_Z(x)\cr h_W(x)},\eee
and obtain\bb
V(\varphi(x))=V(v)+\frac{1}{2}m_H^2H^2(x)+{\rm
terms\ of \ order\ 3\ and\ 4\  in}\
H(x),\ h_Z(x),\ h_W(x),\eee
 with
\bb m_H=\sqrt 2\,\mu.\eee
We come back to the first CL example of section 6.
If we write $\omega=1/2\,\omega_{\mu\nu}\de x^\mu
\de x^\nu\in\Omega^2M$ then $\omega*\omega=
1/2\,\omega_{\mu\nu}\omega^{\mu\nu}
\de x^0\de x^1\de x^2\de x^3$. Consider the Yang-Mills
Lagrangian in equation (\ref{Vt}) on Minkowski
space,
\bb-\t[\rho(F)*\rho(F)\,z]=-\frac{1}{2}
\t\left[\rho(F_{\mu\nu})\rho(F^{\mu\nu})\,z\right]\,
\de x^0\de x^1\de x^2\de x^3.\eee
This term is nothing else but
$-1/4\,\left(F_{\mu\nu},F^{\mu\nu}\right)$. Hence
\bb\frac{1}{2}\t\left(\rho(a,b)^*\rho(a',b')\,z\right)
=\frac{1}{2}\t(a^*a')(y_1+y_2)+b\bar b'(y_1+y_2)\eee
is by comparison equal to
\bb\frac{1}{4}\left((a,b),(a',b')\right)=
\frac{1}{4}\left(\frac{2}{g_2^2}\t(a^*a')+
\frac{1}{g_1^2}\bar bb'\right).\eee
Consequently
\bb g_1^2=\frac{1}{2}\frac{1}{(y_1+y_2)},\qq
g_2^2=\frac{1}{(y_1+y_2)},\eee
and $\sin^2\theta_w=1/3$.
The remaining two pieces of the Euclidean Lagrangian
(\ref{Vt})
 read in Minkowski space
\bb 2L\left|(\pa_\mu+\rho(A_\mu))\varphi\right|^2
-K\left(1-|\varphi|^2\right)^2,\eee
and after the proper rescaling of the scalar
\bb
\frac{1}{2}\left|(\pa_\mu+\rho(A_\mu))
\varphi\right|^2-
K+\frac{1}{2}\frac{K}{L}|\varphi|^2
-\frac{K}{16L^2}|\varphi|^4.\eee Comparing with
equation (\ref{Va}) we have
\bb\lambda=\frac{K}{16L^2},\qq
\mu^2=\frac{K}{L}\eee and
\bb
m_W^2=g_2^2\frac{\mu^2}{16\lambda}
=\frac{L}{y_1+y_2},\qq
m_H^2=2\mu^2=2\,\frac{K}{L}.\eee

 \end{document}